\documentclass[fleqn,12pt,twoside]{article}
\usepackage{espcrc1}

\usepackage{graphicx}
\usepackage[figuresright]{rotating}

\newcommand{\AmS}{{\protect\the\textfont2
  A\kern-.1667em\lower.5ex\hbox{M}\kern-.125emS}}

\title{\vskip -.8in {\normalsize Presented at the 18th International IUPAP Conference
on Few-Body Problems in Physics, Santos, Sao Paulo, Brazil, August 21-26, 2006}
\break
\break
Exploring the Spin Structure of the Proton with Two-Body Partonic 
Scattering at RHIC}

\author{J. Sowinski
{\it for the STAR Collaboration}\break
\break
Indiana University Cyclotron Facility and Department of Phyiscs\break
2401 Milo B. Sampson Ln.\break
Bloomington, Indiana 47408\break
sowinski@indiana.edu
}
       
\begin{document}

\maketitle

\begin{abstract}

The STAR collaboration at the Relativistic Heavy Ion Collider is using
polarized proton beams at $\sqrt{s}$ = 200 GeV to study the spin
structure of the proton.  The first results for the double spin
helicity dependence of inclusive jet production are presented along
with projections for additional data taken in 2005 and 2006.  When
fully analyzed these data sets should place strong constraints on the
possible contribution of gluonic spin to the proton spin as expressed
by $\Delta$G.  Future studies using 2-jet or photon-jet coincidences
to map out the gluon spin distribution vs. the gluon's momentum fraction of the
proton are discussed.
\end{abstract}

\section{Introduction}
It would seem that the question of the contribution of 3 valence
quarks to the spin of the nucleon should be a relatively straight
forward 3-body problem.  And in a constituent quark model, the
magnetic moments of the baryon octet are explained\cite{magmoment} by 3
parameters representing the magnetic moments of the up, down and
strange quarks, thus fully accounting for the spin of the nucleon.  Ever
since the first polarized deep inelastic scattering (DIS) measurements
on the proton directly measured the polarization of the quarks, we
have been confronted with a puzzle in that the quarks, as described in
the infinite momentum frame, seem to account for at most 1/3 of the
proton's spin\cite{disreview}.  Of course in perturbative QCD (pQCD)
the picture of the proton is a many-body problem, with a large number
of virtual gluons and sea quarks popping in and out of
existence.  Current efforts are directed at trying to learn about
contributions to the proton's spin from gluon spin and orbital angular
momentum of the quarks and gluons.  DIS measurements are sensitive to
the gluons only through scaling violations, requiring global fits to
the full kinematic range of the measurements and, up to
now\cite{compass}, have not provided strong constraints on the gluon
contribution, $\Delta$G.

At the Relativistic Heavy Ion Collider (RHIC) we are investigating the
spin of gluons in the proton via two-body hard scattering of quarks
and gluons\cite{rhicspinrev}, analogous to what we nuclear physicists
call quasi-free scattering.  The source of the gluons and quarks are
$\sqrt{s}$~=~200 GeV polarized proton collisions in RHIC.  Processes
that are sensitive to the spin of the gluons include simple
scattering, $q + g \rightarrow q + g$ and $g + g \rightarrow g + g$.
The two outgoing partons fragment to produce colorless particles,
mesons and baryons, correlated in space in a cone and known as a jet.
Of course these processes sensitive to the gluon polarization are
mixed with $qq$ and $q\overline{q}$ scattering, partially diluting the
signal of interest.  Sensitivity to the partonic spin is gained
through the longitudinal double spin asymmetry also known as the spin
correlation parameter,
$$A_{LL} = {{\sigma^{++} - \sigma^{+-}}\over{\sigma^{++} + \sigma^{+-}}}$$
where $\sigma^{++}$ refers to the cross section when both beams have
the same helicity and $\sigma^{+-}$ the case where they have opposite
helicity.  This quantity is large for many partonic scattering
processes giving good sensitivity to the spin of the partons involved
in the scattering.  Underlying the formalism are the familiar
assumptions that the scattering is ``free'', that final and initial
state interactions are small or calculable, and that the 2-body cross
section, the spin dependent momentum distributions and the
fragmentation are factorizable, i.e. do not depend on each other.  The
cross sections for inclusive jets are relatively large and have been
the channel first studied at RHIC as the luminosity has developed in
the initial years of polarized proton operation.  The performance of
RHIC in 2006 gives us optimism to proceed in future years with
studies of smaller cross section channels such as quark-gluon Compton
scattering $q + g \rightarrow q + \gamma $, which has some advantages
that will be discussed below.

\section{STAR Measurements}
The inclusive jet measurements I will present were taken with the STAR
detector at RHIC.  STAR sits at one of six interaction
regions in the 3.8 km circumference double ring collider.  The other
interaction regions are populated by Phenix which is the other large
detector and heavily involved in the spin program, Phobos, Brahms,
pp2pp and polarimetry.  Spin runs so far have primarily consisted of two
100 GeV polarized proton beams colliding at the
center of STAR and PHENIX plus other regions as required.  Test runs
up to beam energies of 250 GeV have been made and this will 
eventually be an important
running energy.  Polarization and luminosity have steadily
increased\cite{rhicprog} year to year, with the polarization reaching
60\% compared to a design value of 70\%, and luminosity within a factor of 3
of the design goal for $\sqrt{s}$=200 GeV running in 2006.

STAR\cite{starnim}, which stands for the Solenoidal Tracker at RHIC,
is a large volume solenoidal magnet (B=0.5 T) with axis along the beam
line with most of its volume filled with a time projection chamber
which provides tracking and momentum analysis of charged particles.
The other detector subsystems of interest for the results presented
here are scintillators near the beam line at each end of
the magnet, known as beam-beam counters, used for triggering, local
polarimetry and luminosity monitoring, and Pb-plastic scintillator
electromagnetic (EM) calorimeters covering a range
approximately 
15$^{\circ}$ to 140$^{\circ}$ with respect to the beam line.  These
detectors have essentially complete coverage in $\phi$ giving STAR the
large solid angle coverage necessary to detect the multiple charged
particles and decay photons making up a jet which can span a cone of
radius 60$^{\circ}$, and moreover provides the coincident solid angle
to detect back to back dijets or photon-jet coincidences.

\begin{figure}[tbh]
\begin{minipage}[t]{71mm}
\includegraphics[width=7.7cm,clip]{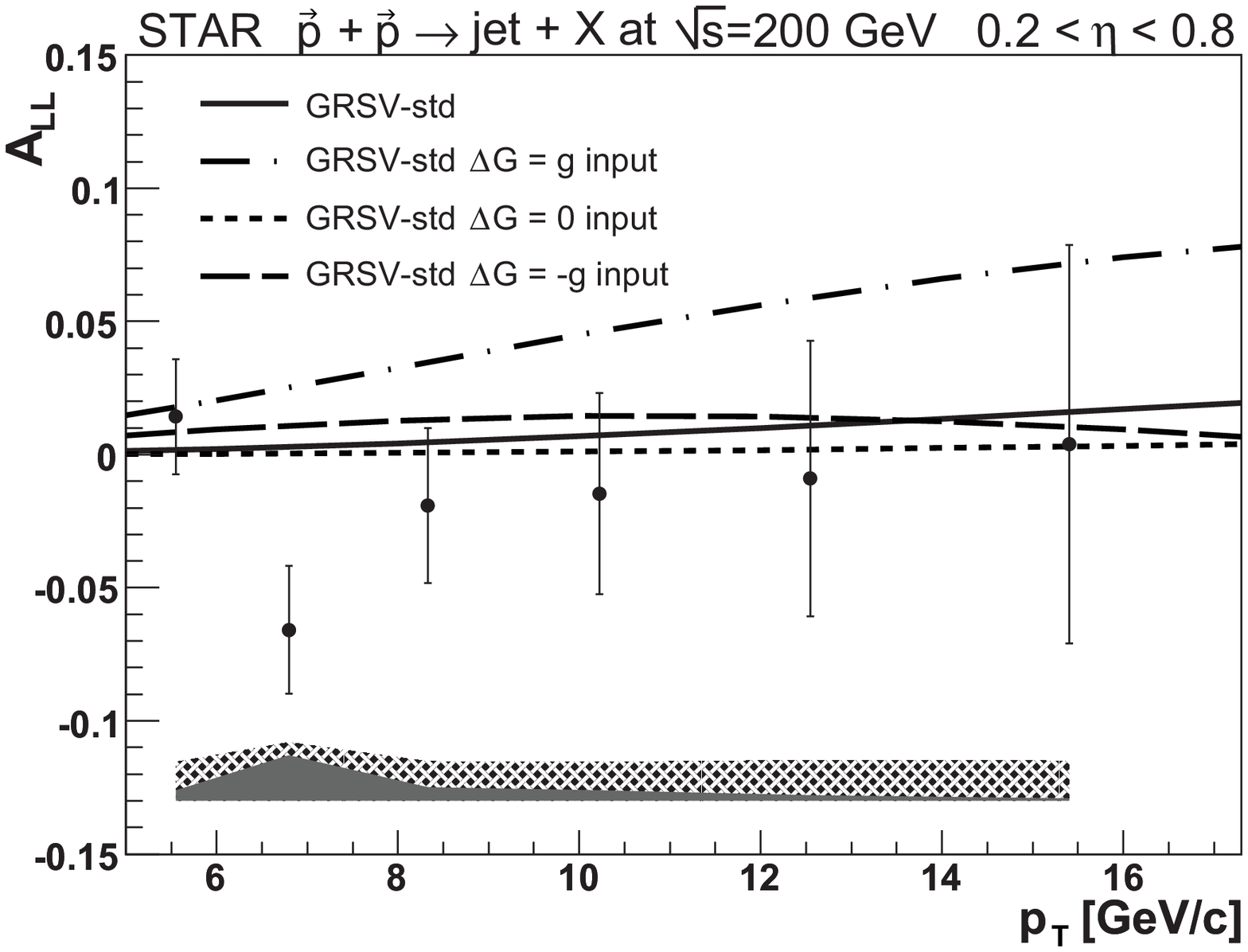}
\label{fig:xsec}
\end{minipage}
%
%
\begin{minipage}[t]{82mm}
\includegraphics[width=7.7cm,clip]{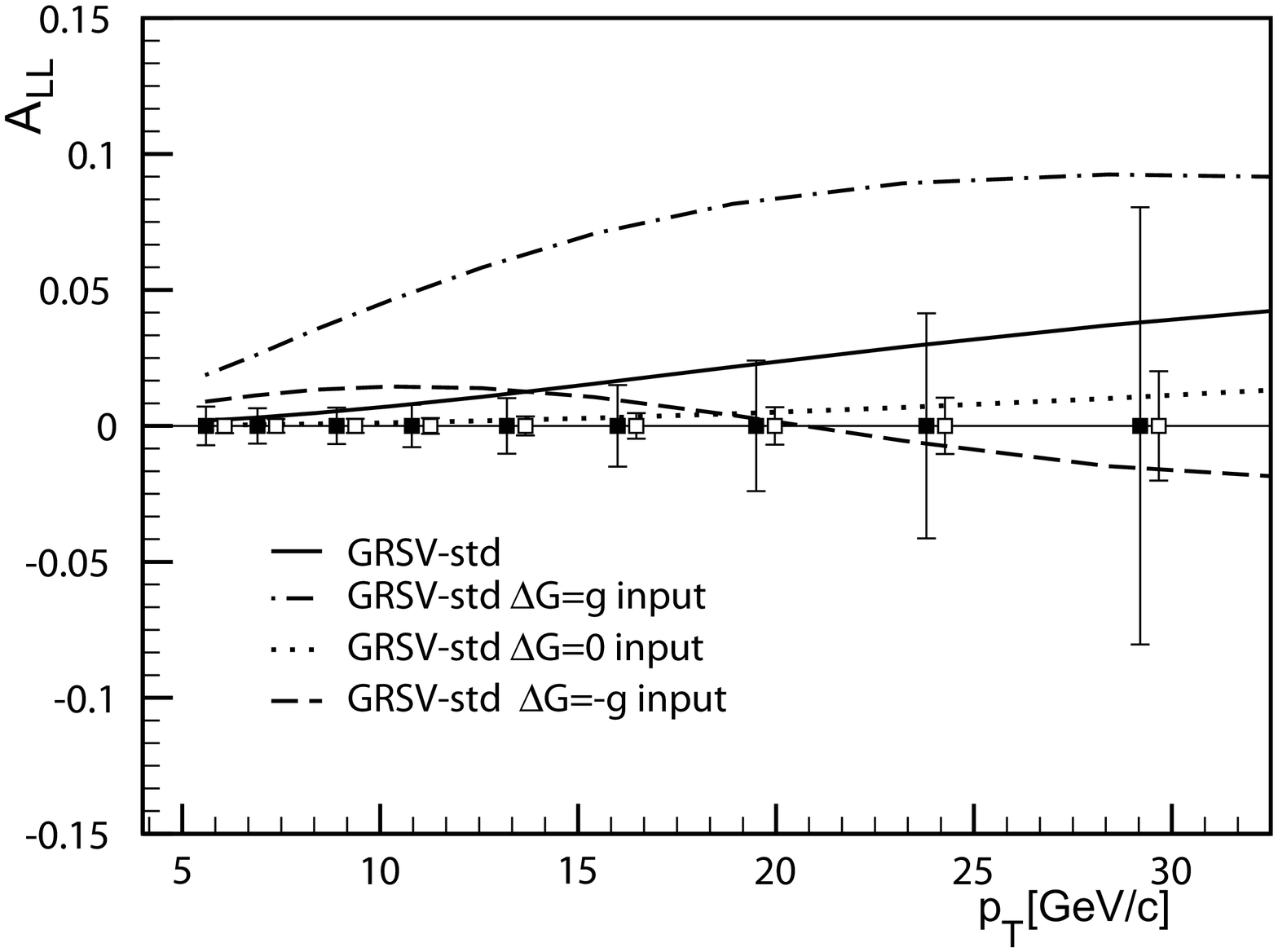}
\label{fig:ALL}
\end{minipage}
Figure 1.  $A_{LL}$ for combined 2003 and 2004 inclusive jet data (left) along with the projected
statistical precision for data taken in 2005 and 2006 (right).

\end{figure}

Jets are readily identified by eye in event displays of large solid
angle detectors such as STAR as groups of tracked charged particles
and photons from decayed mesons in the EM calorimeters grouped in a
cone.  Particle - particle correlations also show the jet nature and
their disappearance in heavy ion collisions has been important in
establishing the new form of matter discovered at RHIC\cite{jetCorr}.
STAR uses the EM component of the jets in the calorimeters to trigger
on candidate jet events.  To do a quantitative comparison between jet data
and pQCD calculations a common jet definition, an algorithm for
finding jets, must be agreed upon.  We are currently using a
midpoint-cone algorithm developed\cite{jetcone} in work at the
Tevatron.  Individual particles above a certain transverse momentum
are used as seeds to establish a cone and the energy in that cone is
collected.  There is a method to merge and split the cones, finally
arriving at a stable definition of jets for the event.  In the
2003-2004 data set, with only about 1/3 of the EM calorimetry
participating, we typically find at most one jet in an event.  In 2006
with full EM calorimetry, events with jets have two jets in the
acceptance over 50\% of the time.  Events with 3 or more jets are also
seen and are expected when hard gluons are radiated during the
partonic scattering or fragmentation.

In the left portion of Fig. 1 are shown our results for the combined
2003 and 2004 longitudinal spin correlation $A_{LL}$ for inclusive
jets recently submitted for publication\cite{jetpap}.  The data are
plotted vs. the magnitude of the momentum transverse to the beam,
p$_T$, which is a measure of the scale the collision probes much as
momentum transfer is used in electron scattering.  The cross section
results extracted from this data compare favorably with next to
leading order pQCD calculations over many orders of magnitude,
particularly in slope.  With the cross section giving
encouragement that pQCD applies to the jets we are seeing, we compare
to next-to-leading-order pQCD calculations\cite{vogel} with different
values of the gluonic contribution to the proton's spin, $\Delta$G.
The different calculations shown, span the range of allowed values,
from fully polarized along the protons helicity, $\Delta$G=g, to
fully anti-aligned, $\Delta$G=-g.  The GRSV std. calculation (solid
line) uses
a polarized gluon distribution typical of best fits to DIS data.  This
first A$_{LL}$ inclusive jet data set rules out only the largest
values of $\Delta$G, and in particular $\Delta$G=g at the 98\% c.l.

In addition to the data presented here we have taken data sets for
inclusive jets in 2005 with improved beam luminosity and polarization
and again in 2006 with further improved beam properties and the full
calorimeter solid angle active and in the trigger.  The size of the
error bars expected for 2005 and 2006 are presented in the right hand
portion of Fig.~2.  The 2005 data will be available by the fall of 2006.

One difficulty with the inclusive jet results is that each data point
averages over a large range of the gluon distribution in Bjorken $x$,
the fraction of the proton's momentum carried by the gluon in the
infinite momentum frame.  For example it would be difficult to
determine whether a small $\Delta$G comes from a value uniformly small
in $x$ or a distribution that changes signs.  By detecting 2 jets or a
gamma and jet in coincidence we have the capability in STAR to map out
the $x$ dependence.  The measured scattering angles are well
correlated with the partonic scattering angles and ratio of the 2
partonic momentum fractions, $x_a/x_b$.  The gamma-jet coincidences
are particularly attractive, despite their smaller cross section, in
that they strongly select for $qg$ interactions.  In addition,
the gamma p$_T$ better preserves the partonic level $p_T$ scale
than the jets, where the energy scale is smeared in fragmentation and
detection.  Exploiting these kinematics allows one to extract the 
individual $x_q$ and $x_g$ in the regime where one of them is 
significantly larger than
the other and thus most likely to be identified with the quark.  
The goal in future running
at 200 GeV is to measure $\Delta$G(x) for 0.03$\le$x$\le$0.3.
Additional running at 500 GeV should allow us to extend to lower $x$
values and provide important cross checks where the data sets overlap.

\section{Summary}
RHIC and STAR have made a lot of progress over the past 6 years and we
are starting to contribute in a significant way to an accounting
of the proton's spin.  Data already taken should place significant
constraints on the magnitude of the gluon's contribution to the spin
of the proton.  There are other measurements we are performing, with
transversely polarized beams for example, I have not had time to
discuss which are making important contributions\cite{morestar} to
other aspects of the defining the proton's structure as well.  We
look forward to the future measurements which have the capability to
go beyond broadly averaged quantities to true distributions of the
gluon spin.

\end{document}